\documentclass[11pt,twocolumn]{article}

\usepackage{fullpage}
\usepackage{amsfonts,amssymb,amsmath}
\usepackage{tikz,pgfplots}
\usepackage{graphicx}
\usepackage{float}
\usepackage{subcaption}
\usepackage{hyperref} 
\usepackage{multicol}
\usepackage[utf8]{inputenc}
\usepackage{cite}
\usepackage{authblk}

\begin{document}

	\title{Intensity-Based Criterion for Determining Exceptional Point in Parity-Time (PT) Symmetric Coupled Array of Optical Waveguides}

	\author[1,2]{Mahla~Bahar}
	\author[1,*] {Mojtaba~Golshani  }
	\author[3]{Mostafa~Motamedifar}
	\author[4]{Khatereh~Jafari}

	\affil[1]{Department of interdisciplinary Physics and Technology, Shahid Bahonar University of Kerman, Kerman, 7616913439, Iran}
	\affil[2]{Faculty of Physics, Isfahan University, Isfahan, 8174673441, Iran}
	\affil[3]{Department of Physics, Shahid Bahonar University of Kerman, Kerman, 7616913439, Iran}
	\affil[4]{Center for Quantum Engineering and Photonics Technologies, Sharif University of Technology, Tehran, Iran}
	\affil[*]{Corresponding author: golshani@uk.ac.ir}
	
	\date{}
	

	\maketitle
	
	\twocolumn[
	
	\begin{abstract}
		In this study, we investigated the propagation pattern and the site-to-site correlation function in a PT-symmetric waveguide array with different input quantum states. 
		Recognizing the stark difference in propagation pattern before and after the PT symmetry-breaking point, we have developed a novel, straightforward intensity-based criterion to determine the exceptional point (EP).
		This new criterion shows excellent agreement with those obtained by directly computing the Hamiltonian's eigenvalues.
		Given the computational complexity of Hamiltonian diagonalization, our proposed criterion provides a highly efficient and valuable alternative for identifying the PT symmetry-breaking point.
		Importantly, the proposed criterion is not restricted to the specific system studied here, but is generally applicable to a wide class of systems that can be described within the tight-binding framework.
		\\
		\\
		\\
		
	\end{abstract}
	]
	

	\section{Introduction}
	
	Over the past decade, discrete optical systems-particularly waveguide arrays-have played a crucial role in controlling the behavior of light~\cite{szameit2010discrete}. Optical waveguide arrays consist of several single-mode waveguides arranged at specific distances. In this system, light is transferred between adjacent waveguides through evanescent coupling. The most effective method for manufacturing these waveguides is direct writing with femtosecond lasers~\cite{liu2019femtosecond}.
	Optical waveguide arrays represent one of the most advanced platforms for quantum information processing and quantum system simulations. This pivotal role is widely substantiated by extensive research, including experimental boson sampling~\cite{tillmann2013experimental}, spatial entanglement in photonic lattices~\cite{di2013einstein,rai2010quantum}, quantum walks of correlated photon pairs~\cite{poulios2014quantum}, generation of high-order single-photon W-states~\cite{grafe2014chip}, experimental observation of N00N state Bloch oscillations~\cite{lebugle2015experimental}, two-photon correlation and transport in disordered lattices~\cite{xu2015two} and anomalous photon transparency in non-Markovian coupled waveguides~\cite{silva2025anomalous}.
	\\
	The ability to control various parameters within waveguides offers a significant advantage for investigating PT-symmetric systems~\cite{golshani2014impact,biesenthal2019experimental,klauck2019observation,KHAZAEINEZHAD2017387,KHAZAEINEZHAD2013299,KHAZAEINEZHAD201339,PhysRevA.87.033817,szameit2010discrete}. PT-symmetry in quantum mechanics was first explored in 1998, following the discovery of Hamiltonians that remain invariant under both parity and time-reversal operations~\cite{praveena2023review}. Notably, these non-Hermitian Hamiltonians can exhibit entirely real spectra~\cite{RevModPhys.96.045002}. PT-symmetry was introduced into optics due to the similarity between the paraxial wave equation and the Schrödinger equation, with the refractive index serving as the optical equivalent of the potential~\cite{du2022light}. A PT-symmetric system in optics can be realized by adjusting the real and imaginary components of the refractive index~\cite{bendix2010optical}. For PT-symmetry to hold in optics, the real part of the refractive index needs to be spatially even, while its imaginary part must be odd function of position \cite{ozdemir2019parity}. Since the imaginary part of the refractive index defines gain and loss, a PT-symmetric system can be achieved in optics by tuning these parameters accordingly. In a two-component optical system, real eigenvalues arise when the gain in one component compensates for the loss in the other. In contrast, in the weak coupling regime, if the gain from one component cannot fully compensate for the loss in the other, the system exhibits complex eigenvalues. This phase transition from real to complex spectra occurs at the EP~\cite{wiersig2020review,klaiman2008visualization}.\\
	A non-Hermitian Hamiltonian typically leads to complex eigenvalues, where the negative imaginary part represents a damping constant associated with the finite spectral linewidth of the resonance energy \cite{wang2023non}. The real part of the eigenvalue determines this linewidth. In classical optical systems, the resonance frequencies and modes correspond to the eigenvalues and eigenstates of the Hamiltonian. Non-Hermiticity can arise due to absorption, gain, and radiation. When the Hamiltonian is non-Hermitian, i.e $ \hat{H} \neq \hat{H}^\dagger $, its eigenstates lose their orthogonality with respect to the inner product used in the standard (Hermitian) formalism of quantum mechanics. However, this orthogonality can be restored by employing a new definition for the inner product, known as the c-product~\cite{moiseyev2011non}. The non-orthogonality, under the standard inner product of quantum mechanics, is maximized at the EP, where the eigenstates of the Hamiltonian coalesce. In this scenario, the states become parallel to one another. These points are considered singularities within non-Hermitian systems. Near an EP, one or more eigenvalues and their corresponding eigenstates coalesce, leading to a phase transition. This transition, known as a PT-symmetry phase transition, results in the emergence of complex eigenvalues~\cite{wiersig2020review,klauck2025crossing}.\\ 
	To compute the EP, the eigenvalues of the system must be calculated, then, the point at which one or more eigenvalues become complex is identified as the PT-symmetry breaking point or EP. Alternatively, the EP can be determined from the propagation pattern. Due to the exceptional growth that occurs after the EP, when eigenvalues turn complex, the propagation pattern exhibits a distinct difference before and after the EP. This approach provides a practical means of defining the EP.\\
	A particularly intriguing application of PT-symmetry breaking points in photonics is in sensors, as systems near the EP exhibit a significantly enhanced response to even minor external disturbances. As the system approaches the n-th order Exceptional Point ($EP_{n}$), n eigenvalues and their corresponding eigenstates merge. The primary distinction between EPs and conventional degeneracies, such as Diabolic Points (DPs), lies in the enhanced sensitivity of the system to external disturbances~\cite{wiersig2020review}.\\
	Another notable application in PT-symmetric optical systems is the achievement of zero group velocity at an exceptional point, where two optical modes coalesce. This effect, tunable across a wide range of frequencies and bandwidths, has been demonstrated in coupled waveguides with balanced gain and loss~\cite{goldzak2018light}.\\
	Consequently, identifying the EP in optical waveguide arrays, and investigating the propagation patterns across the PT-symmetry breaking threshold is of critical importance. EPs play a fundamental role in PT-symmetric systems. As spectral singularities in open quantum systems, they exhibit an enhanced sensitivity to small external perturbations, making them promising candidates for advanced sensor technologies~\cite{wiersig2020review}.\\
	Thompson et al.~\cite{thompson2010anderson} examined intensity and site-to-site correlations in disordered optical waveguide arrays under Anderson localization, using non-classical (squeezed) and classical-like (coherent) light. However, their system did not exhibit PT symmetry. In 2018, the PT-symmetry breaking point of two curved, closely spaced PT-symmetric waveguides, characterized by a small period, was numerically computed using COMSOL Multiphysics software~\cite{zhang2018parity}. The investigation of boundary effects on propagation dynamics and site-to-site correlations within optical waveguide arrays in the quantum regime was reported in 2023~\cite{rai2010quantum}. In~\cite{klauck2025crossing}, the behavior of two-photon states in non-Hermitian systems across the EP was studied. The EP was identified by analyzing variations in the Hong-Ou-Mandel interference with two indistinguishable input photons. The objective of the present work is to compute the EP in a PT-symmetric optical waveguide array with a quantum input state. To this end, two methods are employed: one based on the analysis of the system’s eigenvalues, and the other through the examination of the propagation patterns.
	Furthermore, we show that the proposed intensity-based criterion is in excellent quantitative agreement with the conventional eigenvalue-based determination of the PT-symmetry breaking point. Considering that the direct diagonalization of the Hamiltonian, especially in quantum regimes characterized by exponentially growing Hilbert spaces, is computationally intensive, the present approach offers a highly efficient and scalable alternative. Notably, the scope of this framework extends well beyond the specific photonic system studied here, encompassing a wide range of Schrödinger-type dynamics that admit a tight-binding representation, such as paraxial wave propagation in optical and mechanical settings, quantum evolution in spin chains, population transfer in multilevel configurations, and condensed-matter systems.
	\\
	The organization of the paper is as follows. In Section~\ref{sec2}, the theoretical background is discussed in detail. Section~\ref{sec3} contains the numerical results and their analysis. A summary and concluding are given in Section~\ref{sec4}.

	\label{sec1}
	\section{Theory}
\begin{figure}
		\centering
		\begin{subfigure}{0.8\linewidth}
			\centering
			\includegraphics[width=1.1\linewidth]{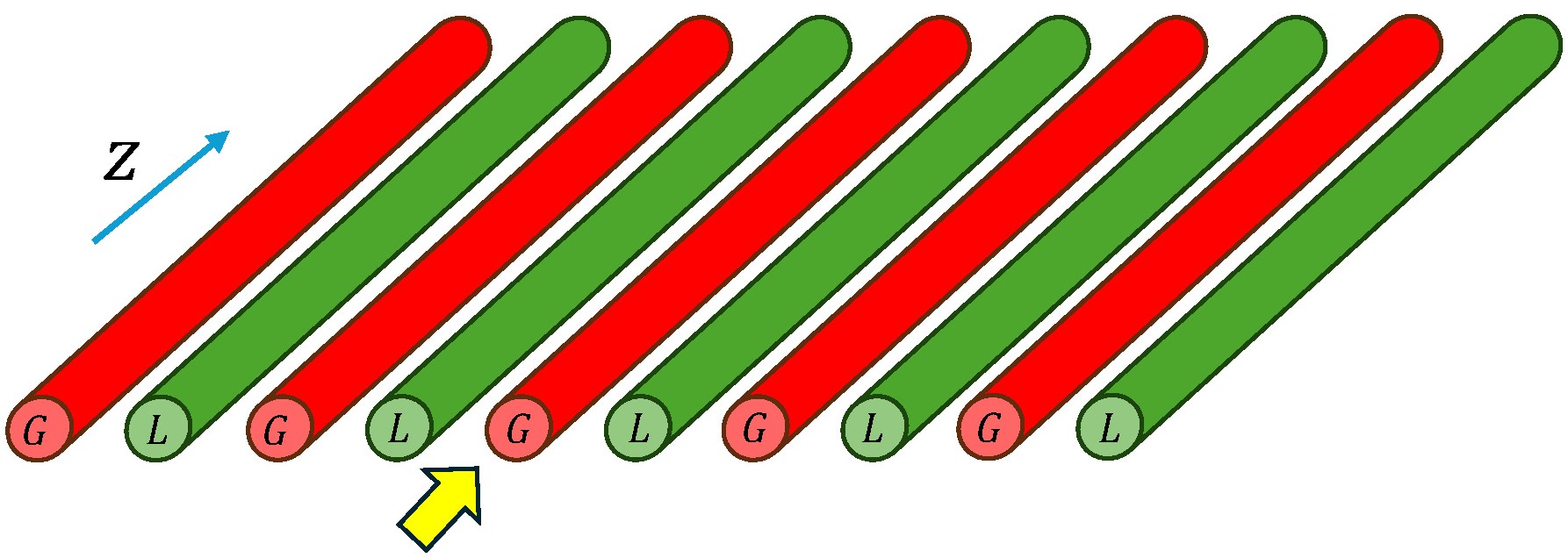}  
			\caption{}
			\label{fig:f1a}
		\end{subfigure}
		
		\vspace{0.5cm} 
		
		\begin{subfigure}{0.8\linewidth}
			\centering
			\includegraphics[width=1.1\linewidth]{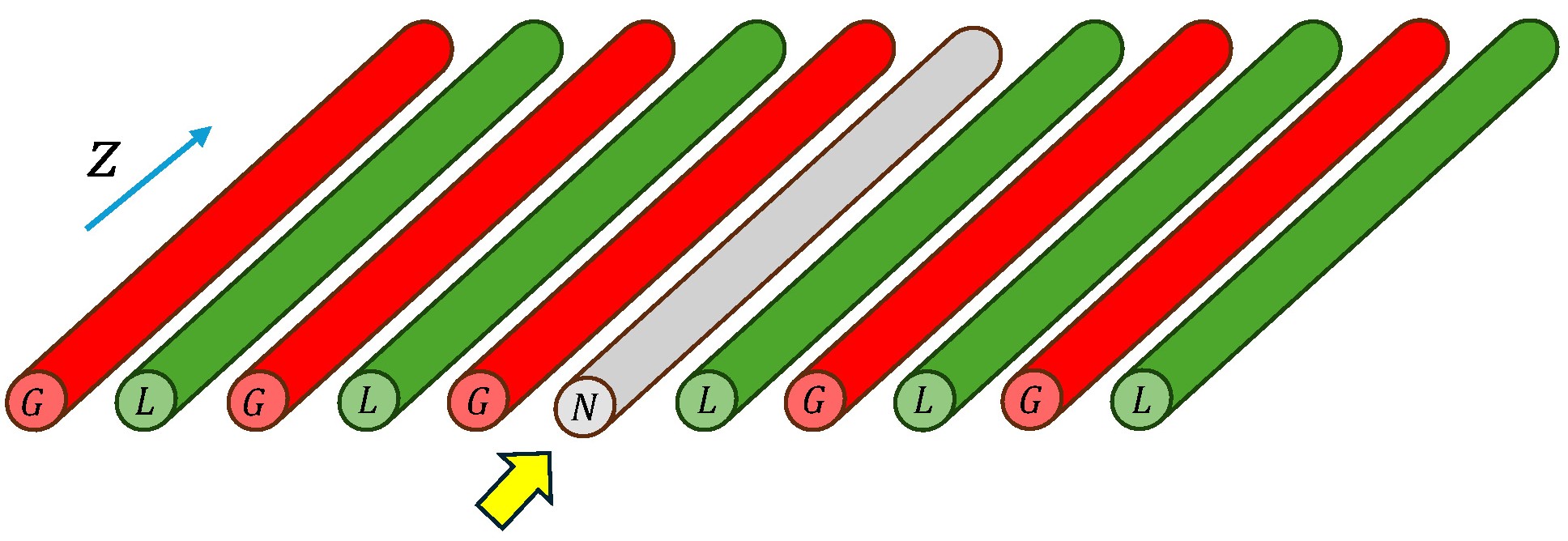}  
			\caption{}
			\label{fig:f1b}
		\end{subfigure}
		
		\caption{PT-symmetric system for (a) even number, and (b) odd, number of waveguides. The waveguides exhibit cylindrical symmetry and are located at equal distances. The colors red, green, and gray correspond to gain (G), loss (L), and neutral (N) conditions, respectively. In these figures, the yellow arrow shows how light enters the system.
		}
		\label{fig:f1}
	\end{figure}

	The system being studied is an optical waveguide array containing N single-mode waveguides.
	This work focuses on analyzing the evolution of a quantum input state in a non-Hermitian PT-symmetric system. 
	Therefore, to satisfy the PT-symmetry criteria, for an even number of waveguides (Figure~\ref{fig:f1a}), the waveguides alternate between loss and gain. 
	In addition, for an odd number of waveguides (Figure~\ref{fig:f1b}), the central waveguide exhibits neither loss nor gain, while the waveguides on either side alternate between loss and gain.\\
	To analyze the propagation of a quantum input state through an optical waveguide array, it is essential to perform the calculations within the framework of quantum optics. Within this regime, the field amplitude is replaced by the creation and annihilation operators, which are used to describe the quantum properties of the field. Let $\hat{a}_j$ and $\hat{a}^\dagger_j$ denote the annihilation and creation operators for the j-th waveguide, respectively. The commutation relations can be expressed as follows \cite{majumder2023effect}:\\
	\begin{equation}
		[\hat{a}_j, \hat{a}_k^\dagger] = \delta_{jk},\quad
		[\hat{a}_j, \hat{a}_k] = 0,\quad
		[\hat{a}_j^\dagger, \hat{a}_k^\dagger] = 0
		\label{eq1}.
	\end{equation}
	If the propagation constant of the $j$-th waveguide is denoted by $\beta_j$ and the coupling coefficient between two adjacent waveguides is $c$ then, in the tight-binding approximation, the Hamiltonian describing the waveguide array is given by \cite{majumder2023effect}:\\
	\begin{equation}
		\hat{H} = \sum_j c \left( \hat{a}_{j+1}^\dagger \hat{a}_j + \hat{a}_j^\dagger \hat{a}_{j+1} \right) + \sum_j \beta_j \hat{a}_j^\dagger \hat{a}_j,
		\label{eq2}
	\end{equation}
	
	where $c$ is the coupling coefficient between nearest-neighbor waveguides, assumed to be identical for all pairs. Moreover, the coupling between higher-order neighboring waveguides is neglected. The first term in equation~(\ref{eq2}) represents the transition of a photon from waveguide $j$ to $j+1$, while the second term corresponds to the reverse process. The final term represents the energy corresponding to each waveguide \cite{majumder2023effect}.\\
	It is important to note that the geometrical configuration of waveguide arrays used in reported experiments explicitly confirms the necessary conditions for applying the tight-binding approximation and evanescent nearest-neighbor coupling. The distance between the centers of adjacent waveguides is typically significantly larger than the excitation wavelength. Therefore, the coupling between adjacent waveguides occurs solely through the overlap of evanescent fields. Consequently, the physics of these evanescently coupled waveguide structures can be accurately described by the tight-binding approximation. This applicability is further validated by the high degree of similarity demonstrated in numerous studies between theoretical predictions (based on the tight-binding model) and experimental results~\cite{tillmann2013experimental,di2013einstein,poulios2014quantum,grafe2014chip,lebugle2015experimental,xu2015two,silva2025anomalous,golshani2014impact,biesenthal2019experimental,klauck2019observation,szameit2010discrete}.
	\\
	In equation~(\ref{eq2}), the propagation constant is given by $\beta_j=\beta_0-ig$ for waveguides exhibiting gain and by $\beta_j=\beta_0+ig$ for those with loss, where $g$ quantifies the gain and loss, and $\beta_0$ is the reference propagation constant that can be removed through normalization  ($\beta_0=0$).\\
	To analyze the evolution of the field operator $\hat{a}_j$, for the $j$-th waveguide, the Heisenberg equation is used \cite{silva2022nonclassical}:
	\begin{equation}
		i \frac{\partial \hat{a}_j}{\partial z} = -[\hat{H}, \hat{a}_j].
		\label{eq3}
	\end{equation}
	It should be noted that the ideal treatment of gain and loss effects in a full quantum optical framework involves the Heisenberg-Langevin equations~\cite{grafe2013correlations} (in the Heisenberg picture) or solving the Master equation~\cite{rai2010quantum,dast2014quantum} (in the Schrödinger picture). However, for simplicity, our study, following the methodology of references~\cite{xu2015two,joglekar2013optical,xu2021dynamical,zloshchastiev2014comparison,echeverri2018comparative,pevrina2019nonclassical,cao2020reservoir},
	adopts the standard non-Hermitian Hamiltonian formalism, where gain and loss are incorporated as imaginary potential terms. The non-Hermitian Hamiltonian captures the average, or mean-field, effect of the environment on the subsystem's dynamics. While it does not account for quantum noise and fluctuations in the same way the full Heisenberg-Langevin or Master equation does, it has been shown to provide an excellent approximation for studying phenomena that are primarily determined by the spectrum of the Hamiltonian, such as exceptional points and their associated dynamics. This approach, which is extensively used and validated in numerous publications on PT-symmetric optics and waveguide arrays (Refs.~\cite{xu2015two,joglekar2013optical,xu2021dynamical,zloshchastiev2014comparison,echeverri2018comparative,pevrina2019nonclassical,cao2020reservoir}), providing a balance between theoretical rigor and computational feasibility.
	\\
	By substituting Hamiltonian~(\ref{eq2}) into equation~(\ref{eq3}), the resulting operator equation is obtained as follows:
	\begin{equation}
		i \frac{\partial \hat{a}_j(z)}{\partial z} = c\, \hat{a}_{j-1}(z) + c\, \hat{a}_{j+1}(z) + \beta_j \hat{a}_j(z),
		\label{eq4}
	\end{equation}
	where $j=1,2,\dots,N$. Since equation~(\ref{eq4}) is linear, there exists a linear relationship between the annihilation operator at propagation length  $z$, denoted $\hat{a}_j(z)$, and the annihilation operator at the initial point ($z=0$), denoted $\hat{a}_m$~\cite{thompson2010anderson}:
	\begin{equation}
		\hat{a}_j(z) = \sum_{m=1}^{N} G_{jm}(z) \hat{a}_m
		\label{eq5}
	\end{equation}
	In this relation $G_{jm}(z)$, known as the Green's function, connects the field at propagation length $z$ in the $j$-th waveguide to the initial field in the $m$-th waveguide. By substituting equation~(\ref{eq5}) into equation~(\ref{eq4}), the following equation is obtained to calculate $G_{jm}(z)$ ;
	\begin{align}
		i \frac{\partial G_{jm}(z)}{\partial z} &= c\, G_{j-1,m}(z) + c\, G_{j+1,m}(z) \nonumber \\
		&\quad + \beta_j G_{j,m}(z)
		\label{eq6}
	\end{align}
	This system of equations consists of $N^2$ scalar differential equations that can be solved numerically to obtain the matrix elements of the Green's function. According to equation (\ref{eq5}), at $z=0$, the initial condition for solving these equations is $G_{jm}(z=0)=\delta_{jm}$.\\
	To analyze the propagation pattern, one must calculate the intensity in each $j$-th waveguide as a function of the propagation distance $z$, which corresponds to the expectation value of the photon number in that waveguide.
	\begin{equation}
		I_j(z) = \langle \hat{a}_j^\dagger(z) \hat{a}_j(z) \rangle
		\label{eq7}
	\end{equation}
	Substituting equation~(\ref{eq5}) into equation~(\ref{eq7}) yields the following expression for the intensity:
	\begin{equation}
		I_j(z) = \sum_{m,n} G_{jm}^*(z)\, G_{jn}(z) \, \langle \hat{a}_m^\dagger \hat{a}_n \rangle
		\label{eq8}
	\end{equation}
	The resulting expression represents the probability of detecting photons in the $j$-th waveguide at a propagation distance of $z$.\\
	Intensity, provides information about the probability of detecting photon at a specific location. This metric is useful for understanding how optical energy is distributed within the array. However, intensity alone cannot reveal deeper quantum aspects of the system, such as spatial entanglement or photon statistics. In essence, intensity shows the average collective behavior of photons.
	\\
	The site-to-site correlation function is utilized to determine the probability of simultaneous photon detection in waveguides $i$ and $j$:
	\begin{equation}
		g_{ji}^{(2)}(z) = \langle \hat{a}_j^\dagger(z)\, \hat{a}_i^\dagger(z)\, \hat{a}_i(z)\, \hat{a}_j(z) \rangle
		\label{eq9}
	\end{equation}
	Substituting equation~(\ref{eq5}) into relation~(\ref{eq9}) yields the following expression for the site-to-site correlation:
	\begin{align}
		g_{ji}^{(2)}(z) &= \sum_{m,n,k,l} G_{jm}^*(z)\, G_{in}^*(z) \nonumber \\
		&\quad \times G_{ik}(z)\, G_{jl}(z)\, \langle \hat{a}_m^\dagger \hat{a}_n^\dagger \hat{a}_k \hat{a}_l \rangle.
		\label{eq10}
	\end{align}
	The second-order correlation function can reveal whether photons behave randomly, exhibit photon bunching, or photon antibunching. This information is essential for understanding the quantum nature of light and investigating spatial entanglement between photons, which is not observable solely from intensity~\cite{di2013einstein,poulios2014quantum,xu2015two}.
	Photons can travel between distant waveguides through a series of sequential nearest-neighbor hops, thereby causing the optical fields at non-adjacent locations to become correlated. Therefore, while photons are not directly coupled between distant waveguides, they can interact indirectly, via intermediate waveguides, leading to correlations in their detection. The site-to-site correlation function precisely measures this spatial correlation.
	\\
	Relations~(\ref{eq8}) and (\ref{eq10}) show that the intensity and site-to-site correlation, can be calculated at any point along the propagation path if the Green’s function and initial state are known.\\
	A primary aim of this work is to investigate how the propagation pattern and site-to-site correlations evolve for various quantum input states within the system.
	For this purpose, two categories of different input states have been investigated. The first category includes the number state  $|m\rangle_p$, coherent state  $|\alpha\rangle_p$ and squeezed state $|\xi\rangle_p$ input into a waveguide number $p$ (Please note that state $|\psi \rangle_p$ is a shorthand for exact state $|0 \rangle_1|0 \rangle_2  \dots|\psi \rangle_p \dots|0 \rangle_N$, where waveguides with zero input photons are not displayed). The second category is entangled NOON state input into two waveguides, $p$ and $q$, $\frac{  |n\rangle_p |0\rangle_q + |0\rangle_p |n\rangle_q  }{ \sqrt{2} }$. In the latter scenario, two cases of the light input in the middle and boundary waveguides have been investigated. To this end, for different input states, the following expressions are utilized:
	\\
	\\
	\textbf{Number state $|n\rangle_p$,}
	\small
	\begin{equation}
	\begin{aligned}
			&\langle \hat{a}_m^\dagger \hat{a}_n \rangle 
			= n \, \delta_{np} \delta_{mp}, \\[4pt]
			&\langle \hat{a}_m^\dagger \hat{a}_n^\dagger \hat{a}_k \hat{a}_l \rangle 
			= n (n - 1) \, \delta_{mp} \delta_{np} \delta_{kp} \delta_{lp}, \\[4pt]
			&I_j(z) 
			= n\,|G_{jp}(z)|^2, \\[4pt]
			&g_{ji}^{(2)}(z) 
			= n(n-1)\,|G_{jp}(z)G_{ip}(z)|^2.\hspace{70pt}
	\end{aligned}
	\label{eq11}
	\end{equation}
	\\
	\normalsize
	\textbf{Coherent state $|\alpha\rangle_p$,}
	\small
	\begin{equation}
		\begin{aligned}
			&\langle \hat{a}_m^\dagger \hat{a}_n \rangle 
			= |\alpha|^2 \delta_{np} \delta_{mp}, \\
			&\langle \hat{a}_m^\dagger \hat{a}_n^\dagger \hat{a}_k \hat{a}_l \rangle 
			= |\alpha|^4 \delta_{mp} \delta_{np} \delta_{kp} \delta_{lp}, \\
			&I_j(z) 
			= |\alpha|^2 \, |G_{jp}(z)|^2, \\
			&g_{ji}^{(2)}(z) 
			= |\alpha|^4\, |G_{jp}(z) G_{ip}(z)|^2.\hspace{70pt}
		\end{aligned}
		\label{eq12}
	\end{equation}
	\\
	\normalsize
	\textbf{Squeeze state $|\xi\rangle_p$,}
	\small
	\begin{equation}
		\begin{aligned}
			&\langle \hat{a}_m^\dagger \hat{a}_n \rangle 
			= \sinh^2(\xi) \, \delta_{np} \delta_{mp}, \\[5pt]
			&\langle \hat{a}_m^\dagger \hat{a}_n^\dagger \hat{a}_k \hat{a}_l \rangle 
			= \sinh^2(\xi) \left( \sinh^2(\xi) + 2 \right) 
			\delta_{mp} \delta_{np} \delta_{kp} \delta_{lp}, \\[5pt]
			&I_j(z) = \sinh^2(\xi) \, |G_{jp}(z)|^2, \\[5pt]
			&g_{ji}^{(2)}(z) = \sinh^2(\xi) \left(\sinh^2(\xi)+2 \right)
			|G_{jp}(z)G_{ip}(z)|^2.
		\end{aligned}
		\label{eq13}
	\end{equation}
	\\
	\normalsize
	\textbf{Entangled NOON state,}
	\small
	\begin{equation}
		\begin{split}
			&\langle \hat{a}_m^\dagger \hat{a}_n \rangle 
			= \frac{n}{2} \big( \delta_{np} \delta_{mp} + \delta_{nq} \delta_{mp} 
			+ \delta_{np} \delta_{mq} + \delta_{nq} \delta_{mq} \big), \\[6pt]
			&\langle \hat{a}_m^\dagger \hat{a}_n^\dagger \hat{a}_k \hat{a}_l \rangle 
			= \frac{n(n-1)}{2} \bigg(
			\delta_{mp} \delta_{np} \delta_{kp} \delta_{lp} + \delta_{mq} \delta_{nq} \delta_{kq} \delta_{lq} \\
			&\quad + \delta_{n,2} \big( \delta_{mp} \delta_{np} \delta_{kq} \delta_{lq}
			+ \delta_{mq} \delta_{nq} \delta_{kp} \delta_{lp} \big) \bigg), \\[6pt]
			&I_j(z) 
			= \frac{n}{2} \Big(
			|G_{jp}(z)|^2 + |G_{jq}(z)|^2 \\
			&\quad + G^*_{jp}(z) \, G_{jq}(z) + G_{jp}(z) \, G^*_{jq}(z) \Big), \\[6pt]
			&g_{ji}^{(2)}(z) 
			= \frac{n(n{-}1)}{2} \Big(
			|G_{jp}(z)G_{ip}(z)|^2 + |G_{jq}(z)G_{iq}(z)|^2\\
			&\quad + \delta_{n,2} \big( 
			G^*_{jp}(z) G^*_{ip}(z) G_{jq}(z) G_{iq}(z) \\
			&\qquad + G^*_{jq}(z) G^*_{iq}(z) G_{jp}(z) G_{ip}(z) \big) \Big)
		\end{split}
		\label{eq14}
	\end{equation}
	\normalsize
	An additional objective of this study is to determine the EP ($g_c$) of the system. To this end, two different approaches are evaluated. Since the propagation pattern exhibits oscillatory behavior for $g < g_c$ and transitions to an exponential energy growth (in the linear regime) for $g > g_c$, the propagation pattern itself can be used to identify the exceptional point. For this purpose, the following criterion is defined:
	\begin{equation}
		I_r = \exp\left( 
		-\gamma \left| \frac{\bar{I}_2 - \bar{I}_1}{\bar{I}_2 + \bar{I}_1} \right|^\beta 
		\right)
		\label{eq15}
	\end{equation}
	The determination is based on the spatial intensity distribution along the propagation length, with $\bar{I}_1$ and $\bar{I}_2$ representing the mean intensities over all waveguides in the first half ($0 < z < \frac{z_{\text{max}}}{2}$) and second half ($\frac{z_{\text{max}}}{2} < z < z_{\text{max}}$) of the propagation length $z_{\text{max}}$, respectively. Besides, the parameters $\alpha$ and $\beta$ are two positive constants chosen to exhibit a significant change at the EP. 
	In practice, the choice $\gamma = \beta = 10$ provides a robust and reliable characterization of the transition, and the criterion is not highly sensitive to the precise values of these parameters.
	To further clarify the validity and underlying mechanism of the proposed criterion, we examine its behavior in the two distinct regimes of unbroken and broken PT symmetry.
	Before the EP, the energy exhibits oscillatory behavior, so $\bar{I}_1 \simeq \bar{I}_2$. In this case, $\left|\frac{\bar{I}_2 - \bar{I}_1}{\bar{I}_2 + \bar{I}_1}\right|^\beta$ is approximately zero, which results in $I_r = 1$. However, after the EP, as the propagation length increases, the energy grows exponentially, so $\bar{I}_1 \ll \bar{I}_2$. Therefore, $\left|\frac{\bar{I}_2 - \bar{I}_1}{\bar{I}_2 + \bar{I}_1}\right|^\beta$ approaches one, and $I_r = 0$. 
	These distinct behaviors lead to clearly different values of the proposed measure in the two regimes, thereby enabling it to serve as a reliable indicator for identifying the transition point.
	Subsequently, by plotting $I_r$ versus gain/loss factor $g$ and investigating the transition from one to zero in the $I_r$ criterion, the PT-symmetry breaking point $g_c$ can be identified.\\
	A more accurate approach to locate the EP is to diagonalize the system Hamiltonian and calculate its eigenvalues. The EP is found at a critical gain/loss strength $g_c$ where the Hamiltonian's eigenvalues shift from real to complex, which can be observed by plotting their imaginary parts versus $g$. 
	Because of the bosonic statistics of photons, there is no inherent limit to the number of photons residing in each waveguide. Therefore, the Hamiltonian~(\ref{eq2}) is defined in an infinite-dimensional Hilbert space. Consequently, determining its eigenvalues presents a significant analytical and computational challenge.
	Moreover, if the maximum number of photons per waveguide is restricted to $N_p$, the dimension of the aforementioned Hilbert space becomes $(N_p+1)^N$, which exponentially increases with number of waveguides, N.
	The exponential growth of the Hilbert space dimension with the number of waveguides, and the associated computational complexity of its diagonalization, highlight the importance of the criterion introduced in Eq.~(\ref{eq15}) for determining the EP. This is because calculating this criterion does not require Hamiltonian diagonalization; instead, the EP can be determined by solving the $N^2$ coupled linear differential equations in Eq.~(\ref{eq6}) and computing the spatial evolution of the intensity.
	In this research, because the Hamiltonian~(\ref{eq2}) preserves the total photon number, as ensured by the commutation relation $[\hat{H}, \hat{n}_t] = 0$ with $\hat{n}_t = \sum_j \hat{a}_j^\dagger \hat{a}_j$, the system's dynamics can be analyzed within separate photon-number subspaces (zero, one, two photons, etc.), simplifying the problem.\\
	In general, for a waveguide array consisting of $N$ waveguides, the subspace corresponding to $m$ photons forms a Hilbert space of dimension
	$d = \binom{m + N - 1}{m}.$
	The standard basis of this subspace is given by the state vector $| n_1, n_2, n_3, \dots, n_N \rangle = | n_1 \rangle_1 | n_2 \rangle_2 \dots | n_N \rangle_N$, subject to the constraints $n_j\geq 0$ and $\sum_{j=1}^{N} n_j = m$. This state stands for a case with $n_j$ photons at j-th waveguide ($j = 1,2,..,N$). In order to calculate the eigenvalues within this subspace, it is necessary to first determine the elements of the Hamiltonian matrix in the specified basis. Thus, we obtain:
	\small
	\begin{multline}
		\hat{H} \left| \dots , n_j, n_{j+1}, \dots \right\rangle = \sum_{j=1}^{N} \beta_j n_j \left| \dots , n_j, n_{j+1}, \dots \right\rangle \\
		+ c \sum_{j=1}^{N-1} \Big\{ \sqrt{(n_j+1) n_{j+1}} \left| \dots, n_j{+}1, n_{j+1}{-}1, \dots \right\rangle \\
		+ \sqrt{n_j (n_{j+1}+1)} \left| \dots, n_j{-}1, n_{j+1}{+}1, \dots \right\rangle \Big\}
		\label{eq17}
	\end{multline}
	\normalsize
    Therefore, the matrix elements of the Hamiltonian will be as follows:
	\begin{equation}
		\begin{aligned}
			\langle n_1', n_2', n_3', \dots, n_N' | \hat{H} | n_1, n_2, n_3, \dots, n_N \rangle &= \\
			&\hspace{-65mm} \left( \sum_{j=1}^N \beta_j n_j' \right) \delta_{n_1', n_1} \delta_{n_2', n_2} \dots \delta_{n_N', n_N} \\
			&\hspace{-65mm} + c \sum_{j=1}^N \left\{ \sqrt{n_j' (n_{j+1}' + 1)} \delta_{n_1', n_1} \delta_{n_2', n_2} \dots \delta_{n_j', n_j+1} \right. \\
			&\hspace{-65mm} \left. \delta_{n_{j+1}', n_{j+1}-1} \dots \delta_{n_N', n_N} \right. \\
			&\hspace{-65mm} + \sqrt{(n_j' + 1) n_{j+1}'} \delta_{n_1', n_1} \delta_{n_2', n_2} \dots \delta_{n_j', n_j-1} \\
			&\hspace{-65mm} \left.\delta_{n_{j+1}', n_{j+1}+1} \dots \delta_{n_N', n_N} \right\}
		\end{aligned}
		\label{eq18}
	\end{equation}
	
	Equation~(\ref{eq18}) can be used to calculate the eigenvalues of the Hamiltonian in the $m$-photon subspace and to examine their imaginary parts to find the EP. In the next sections, we will calculate and compare the EP of a waveguide array using the two methods discussed.
	\label{sec2}
	\section{Numerical Results}
	
\begin{figure}[!t]
		\centering
		\includegraphics[width=\linewidth]{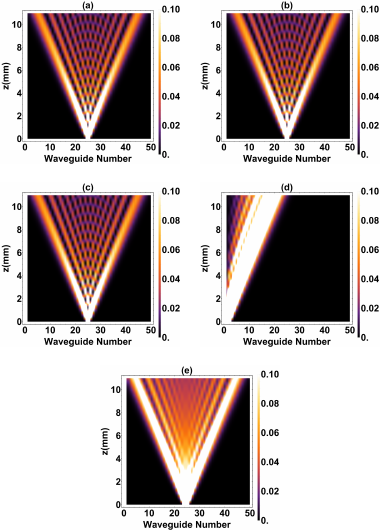}
		\caption{
			Propagation pattern of a waveguide array with $N = 50$ waveguides in the Hermitian regime ($g = 0$) for:  
			(a) number state $|n\rangle_p$ with $n = 1$,  
			(b) coherent state $|\alpha\rangle_p$ with $\alpha^2 = 1$,  
			(c) squeezed state $|\xi\rangle_p$ with $\sinh^2(\xi) = 1$,  
			(d) entangled state between the 1st and 2nd waveguides, and  
			(e) entangled state between the 24th and 25th waveguides.
		}
		\label{fig:fig2}
	\end{figure}

\begin{figure}[!t]
		\centering
		\includegraphics[width=\linewidth]{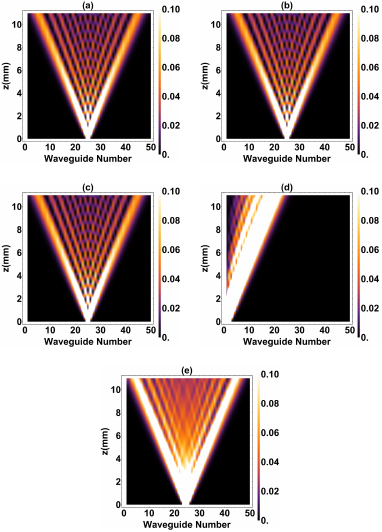}
		\caption{
			Propagation pattern of a waveguide array with $N = 50$ waveguides in the PT-symmetry regime ($g = 0.03$) for:  
			(a) number state $|n\rangle_p$ with $n = 1$,  
			(b) coherent state $|\alpha\rangle_p$ with $\alpha^2 = 1$,  
			(c) squeezed state $|\xi\rangle_p$ with $\sinh^2(\xi) = 1$,  
			(d) entangled state between the 1st and 2nd waveguides, and  
			(e) entangled state between the 24th and 25th waveguides.
		}
		\label{fig:fig3}
	\end{figure}

	Numerical results for the propagation pattern, site-to-site correlation, and exceptional point are presented in this section. As a result, the propagation length is normalized with respect to the coupling constant $c$ ($c=1$). First, the propagation pattern is investigated for a waveguide array consisting of $N=50$ waveguides, with quantum input states including separable and entangled states, in both Hermitian and non-Hermitian systems. Figure~\ref{fig:fig2} shows the results in the Hermitian regime ($g = 0$), for different quantum input states injected into waveguide $p = 25$, including the number state $|n\rangle_p$, coherent state $|\alpha\rangle_p$, and squeezed state $|\xi\rangle_p$, each prepared with an average of one photon ($n = 1$,  $\alpha^2 = 1$, $\sinh^2 \xi = 1$). Additionally, two types of entangled states are considered: $\frac{ \left( |0\rangle_1 |1\rangle_2 + |1\rangle_1 |0\rangle_2 \right) }{ \sqrt{2} }$ and $\frac{ \left( |0\rangle_{24} |1\rangle_{25} + |1\rangle_{24} |0\rangle_{25} \right) }{ \sqrt{2} }$, which are injected into the 1st–2nd and 24th–25th waveguides, respectively. Since the average number of input photons in the separable states is identical, the resulting propagation patterns are equivalent for all three cases, consistent with equations~(\ref{eq11}) to (\ref{eq13}).
	\\
	Moreover, the figures distinctly exhibit the characteristic features of discrete diffraction~\cite{szameit2010discrete}. 
	It is important to note that, in the study of optical waveguides array, two common approaches for handling boundaries are zero-boundary conditions and periodic boundary conditions~\cite{majumder2023effect,joglekar2013optical,xu2021dynamical,Sukhorukov12}.
	In this work, we have employed zero-boundary conditions. Since the primary focus of this paper is not on boundary effects, the propagation length and number of waveguides are chosen such that the light does not reach the boundaries during the propagation. Of course, one could investigate a longer propagation length without boundary effects by increasing the number of waveguides. However, the performed study showed that this does not lead to new results and has no significant effect on the overall propagation pattern. Therefore, for the analysis of the propagation pattern and site-to-site correlation, N=50 waveguides and a maximum propagation length of z=10mm are used.
	Figure~\ref{fig:fig2} demonstrates the propagation pattern within the Hermitian regime, characterized by the absence of gain or loss ($g = 0$). Upon the introduction of gain or loss to the system, and when its magnitude remains below the critical value $g_c$, the system stays in the PT-symmetric regime. Figure~(\ref{fig:fig3}) illustrates that in this regime ($g = 0.03$), the propagation pattern is nearly identical to that in the Hermitian regime. However, unlike in the Hermitian state, the total intensity of light in the PT-symmetric phase is not fixed, instead exhibiting oscillations throughout the propagation due to the effects of gain and loss. For better comparison, in all plots related to the propagation pattern, the total intensity is normalized at each propagation distance.
	It is important to note that Anderson localization, which arises from random disorder, does not occur in our array (with uniform spacing and identical waveguides except for their gain/loss properties)~\cite{thompson2010anderson,majumder2023effect,silva2022nonclassical}. Furthermore, the existence of defect modes, particularly in the central neutral waveguide of figure~\ref{fig:f1b}, is highly dependent on the system's parameters~\cite{PhysRevA.84.013833}. 
	In our model, with uniform coupling across all waveguides, a detailed investigation into their potential existence is beyond the scope of this study. Moreover, the nature of the input excitation is crucial; in our simulations, only one or two waveguides are excited. This input excites a superposition of the system's eigenmodes, including both extended and any potential localized modes, rather than selectively exciting localized ones. Consequently, the observed output pattern is a manifestation of the interference and evolution of all modes. The observed delocalized propagation is therefore a direct consequence of the uniform coupling and the specific excitation scheme, and is consistent with the physics of the system.
		\begin{figure}[!t]
		\centering
		\includegraphics[width=\linewidth]{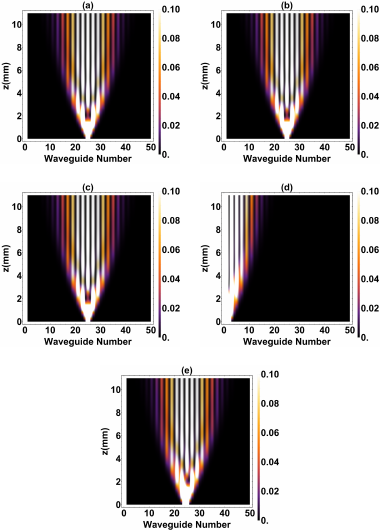}
		\caption{
			Propagation pattern of a waveguide array with $N = 50$ waveguides in the broken PT-symmetry regime ($g = 0.6$) for:  
			(a) number state $|n\rangle_p$ with $n = 1$,  
			(b) coherent state $|\alpha\rangle_p$ with $\alpha^2 = 1$,  
			(c) squeezed state $|\xi\rangle_p$ with $\sinh^2(\xi) = 1$,  
			(d) entangled state between the 1st and 2nd waveguides, and  
			(e) entangled state between the 24th and 25th waveguides.
		}
		\label{fig:fig4}
	\end{figure}
	In Figure~(\ref{fig:fig4}), the propagation pattern of a waveguide array consisting of $N=50$ waveguides with a quantum input state is investigated in the broken PT-symmetry regime $g>g_c$ ($g = 0.6$). As energy grows exponentially in this regime, the waveguides with gain near the input one exhibit much higher intensity than others. Therefore, after normalization, the intensity contributions from the rest of the waveguides are negligible and nearly vanish in the propagation pattern.
	\\
	The study of site-to-site correlations for input number state $|n\rangle_p$ (Eq.~(\ref{eq11})) and NOON state (Eq.~(\ref{eq14})) requires considering input states with more than one photon, as $g_{ji}^{(2)}$ quantifies the probability of simultaneous detection of two photons in distinct waveguides $i$ and $j$. Accordingly, in this work, we compute the site-to-site correlation for different input states with average photon number of two (relations ~(\ref{eq11}) to (\ref{eq14})).
	Figure~(\ref{fig:fig5}) shows the site-to-site correlation for a waveguide array with $N=50$ waveguides at a propagation distance of $z=10$ in the Hermitian regime, where $g=0$.\\
	Apart from a factor, the site-to-site correlation is identical for the number state, coherent state, and squeezed state, with the highest values concentrated near the four corners. This result demonstrates that, in the given state and propagation pattern at \( z = 10 \), the probability of detecting two photons in close proximity (either near waveguide 7 or 43) is approximately identical to the probability of detecting two photons at a specific distance - one near waveguide 7 and the other near waveguide 43.
		\begin{figure}[!t]
		\centering
		\includegraphics[width=\linewidth]{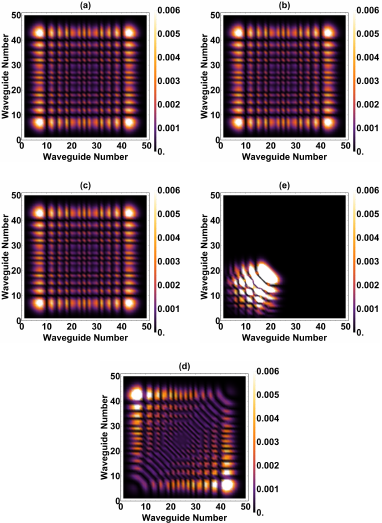}
		\caption{
			Site-to-site correlation a waveguide array with $N = 50$ waveguides in the Hermitian regime ($g = 0$) for:  
			(a) number state $|n\rangle_p$ with $n = 2$,  
			(b) coherent state $|\alpha\rangle_p$ with $\alpha^2 = 2$,  
			(c) squeezed state $|\xi\rangle_p$ with $\sinh^2(\xi) = 2$,  
			(d) entangled state between the 1st and 2nd waveguides, and  
			(e) entangled state between the 24th and 25th waveguides at propagation distance $z=10$.
		}
		\label{fig:fig5}
	\end{figure}
	\begin{figure}[!t]
		\centering
		\includegraphics[width=\linewidth]{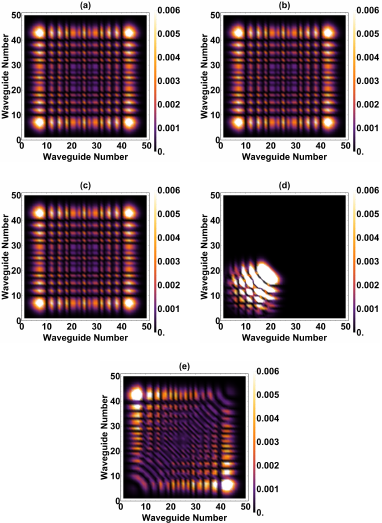}
		\caption{
			Site-to-site correlation a waveguide array with $N = 50$ waveguides in the PT-symmetry regime ($g = 0.03$) for:  
			(a) number state $|n\rangle_p$ with $n = 2$,  
			(b) coherent state $|\alpha\rangle_p$ with $\alpha^2 = 2$,  
			(c) squeezed state $|\xi\rangle_p$ with $\sinh^2(\xi) = 2$,  
			(d) entangled state between the 1st and 2nd waveguides, and  
			(e) entangled state between the 24th and 25th waveguides at propagation distance $z=10$.
		}
		\label{fig:fig6}
	\end{figure}
		\begin{figure}[!t]
		\centering
		\includegraphics[width=\linewidth]{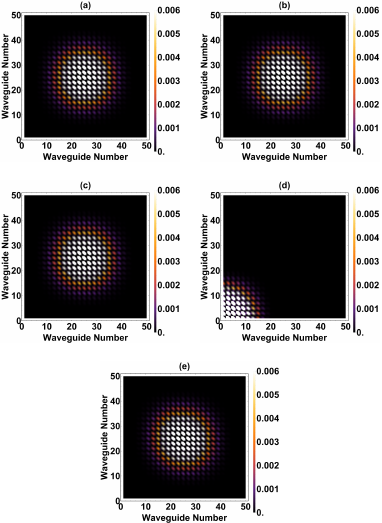}
		\caption{
			Site-to-site correlation a waveguide array with $N = 50$ waveguides in the Broken PT-symmetry regime (g=0.6) for:  
			(a) number state $|n\rangle_p$ with $n = 2$,  
			(b) coherent state $|\alpha\rangle_p$ with $\alpha^2 = 2$,  
			(c) squeezed state $|\xi\rangle_p$ with $\sinh^2(\xi) = 2$,  
			(d) entangled state between the 1st and 2nd waveguides, and  
			(e) entangled state between the 24th and 25th waveguides at propagation distance $z=10$.
		}
		\label{fig:fig7}
	\end{figure}
	The site-to-site correlation shown in Figure~5(e) for an entangled state introduced at the middle of the array indicates that the photons are most probably detected at widely separated positions, approximately at waveguides~7 and~43. This pattern is a signature of spatial anti-bunching. In comparison, Figure~5(d) highlights that entangled photons entering through the array's boundaries tend to remain spatially close, exhibiting a peak near waveguide~20. Therefore, this plot shows spatial photon bunching because the photons exit around a specific waveguide.
	\\
	Similar to Figure~(\ref{fig:fig5}), Figure~(\ref{fig:fig6}) illustrates the case of the unbroken PT-symmetric phase, where the gain/loss parameter satisfies $g < g_c$ ($g = 0.03$). By comparing Figure~(\ref{fig:fig5}) and Figure~(\ref{fig:fig6}), it is evident that the statistical features of the PT-symmetric phase, such as intensity distribution and site-to-site correlation, closely resemble those observed in the Hermitian regime. In other words, the site-to-site correlation exhibits similar behavior regardless of whether the system operates in the PT-symmetric or Hermitian regime. Due to the exact balance between gain and loss, the optical system in the PT-symmetric regime maintains a stable evolution, analogous to the Hermitian case where neither gain nor loss is present.
		
	In Figure~(\ref{fig:fig7}), the site-to-site correlation functions are shown for different quantum input states under the broken PT-symmetry condition, characterized by $g > g_c$ ($g = 0.6$). It is demonstrated that, in the broken PT-symmetric regime, the correlation behavior is modified in contrast to the Hermitian and unbroken PT-symmetric cases. In this regime, the value of the correlation function reaches its maximum near the input waveguide, indicating that the two photons are most likely to be detected in the nearby gain waveguides.\\
	Interference is the physical core behind how photons distribute and how spatial correlation patterns are formed in a coupled quantum system. In systems featuring gain and loss, interference manifests in more complex forms, which indeed include gain-assisted and loss-assisted interference. These phenomena can be revealed in several ways. Gain can amplify specific pathways for photons, while loss attenuates others. This leads to interference patterns that are unevenly influenced by gain and loss, potentially differing significantly from purely Hermitian or passive systems. Moreover, the presence of gain and loss can induce net photon current in specific directions, which is not typically observed in Hermitian systems. In fact, the observed changes in correlation patterns in the presence of gain and loss are directly attributable to these interferences.
	In fact, the observed filamentation (patterned structure) in the site-to-site coherence function is a physical phenomenon resulting from interference and the internal gain/loss within the waveguides.
	\\
	As has been observed thus far, the system exhibits distinct behavior in the PT-symmetric and broken PT-symmetric regimes. Therefore, calculating the EP is essential to more accurately understand the system's behavior.\\
		\begin{figure}[!t]
		\centering
		\includegraphics[width=\linewidth]{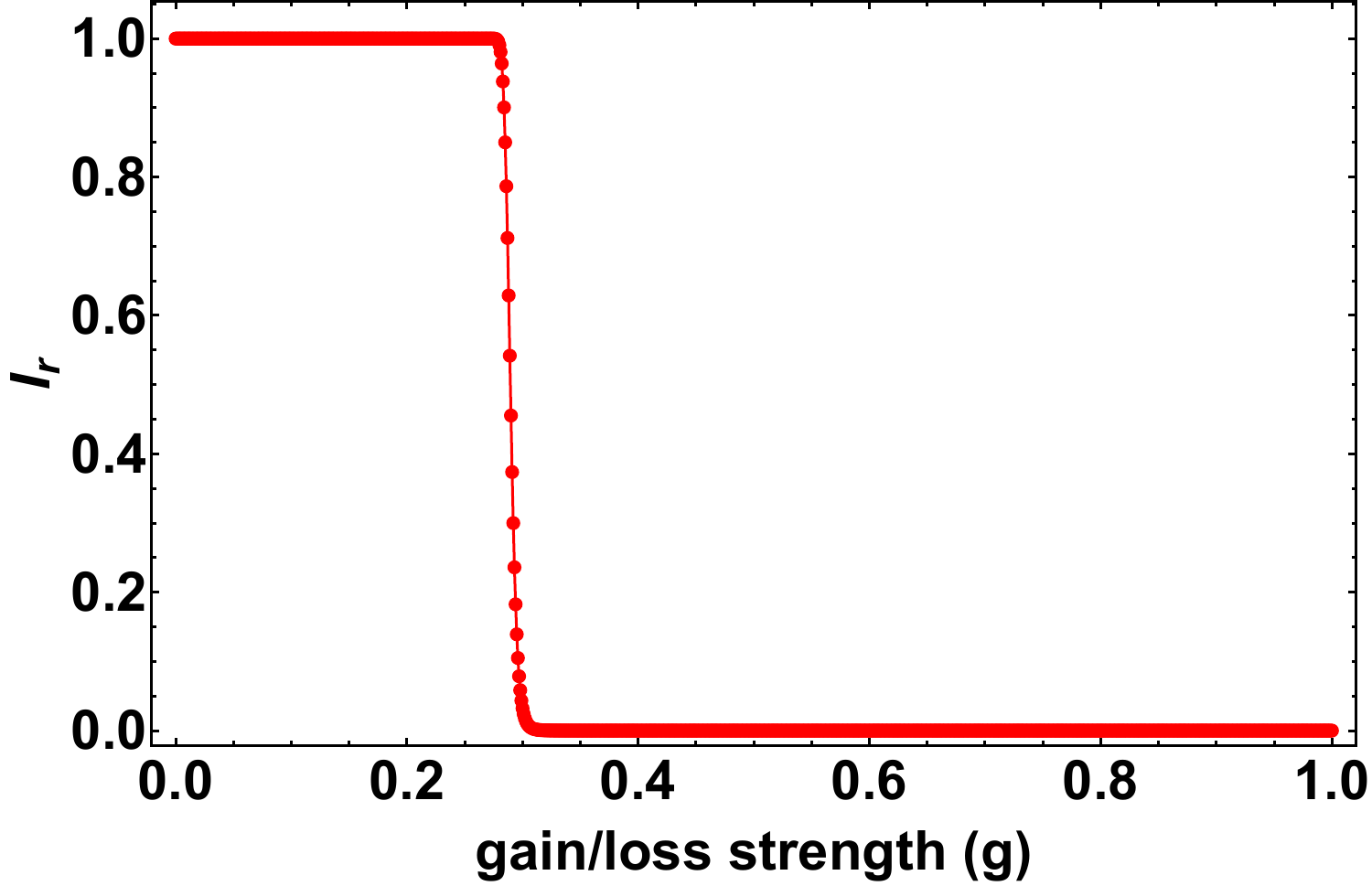}
		\caption{
			$I_r$ as a function of gain/loss parameter $g$ for an entangled input state in the 24th–25th waveguides of a $N=10$ waveguide array.
		}
		\label{fig:fig8}
	\end{figure}
	\begin{figure}
		\centering
		\includegraphics[width=\linewidth]{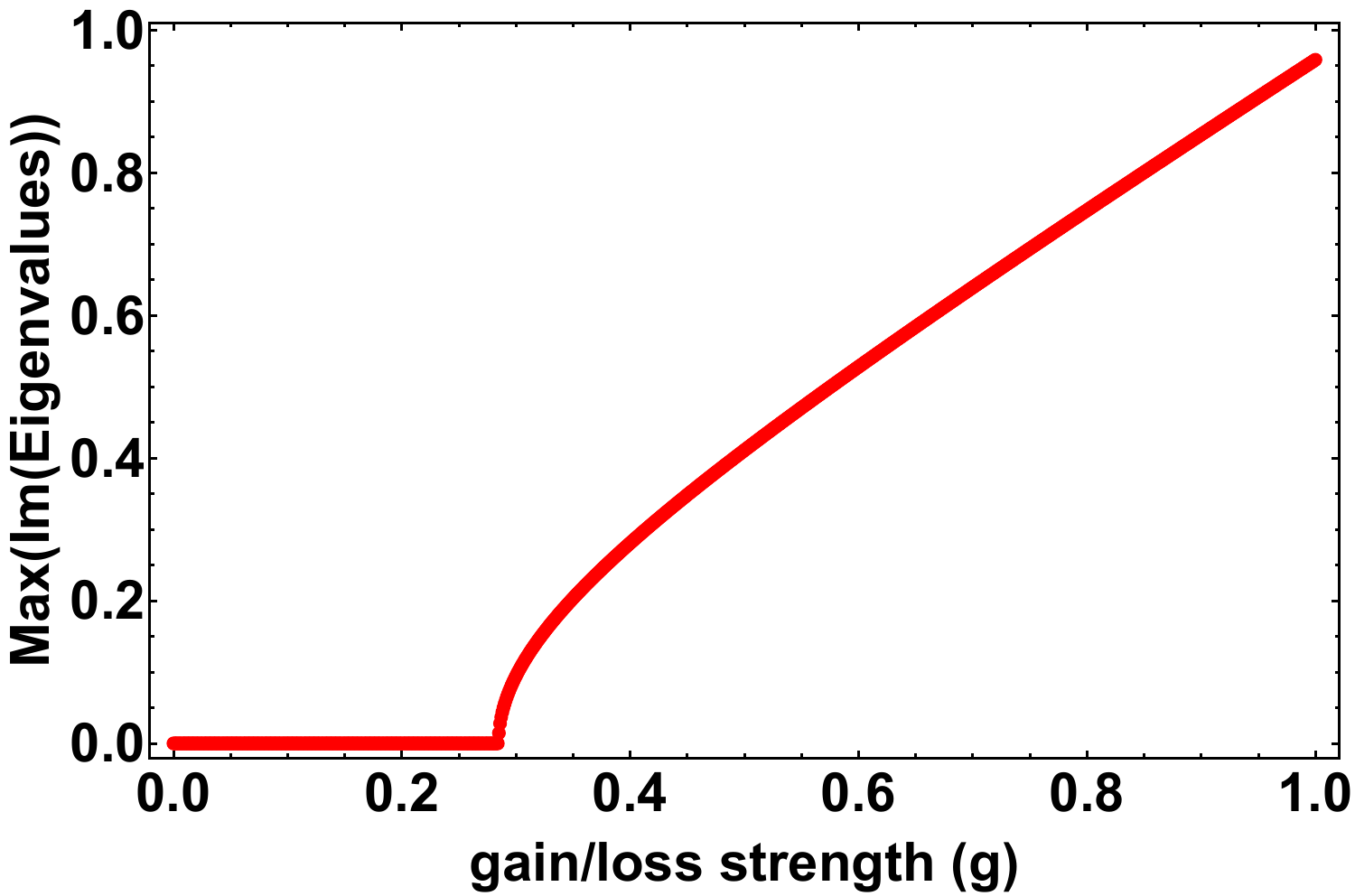}
		\caption{
			Maximum imaginary part of the eigenvalues as a function of the gain/loss parameter $g$ for a waveguide array with $N = 10$ in the single-photon ($m = 1$) subspace.
		}
		\label{fig:fig9}
	\end{figure}
	\begin{figure}[ht]
		\centering
		\includegraphics[width=\linewidth]{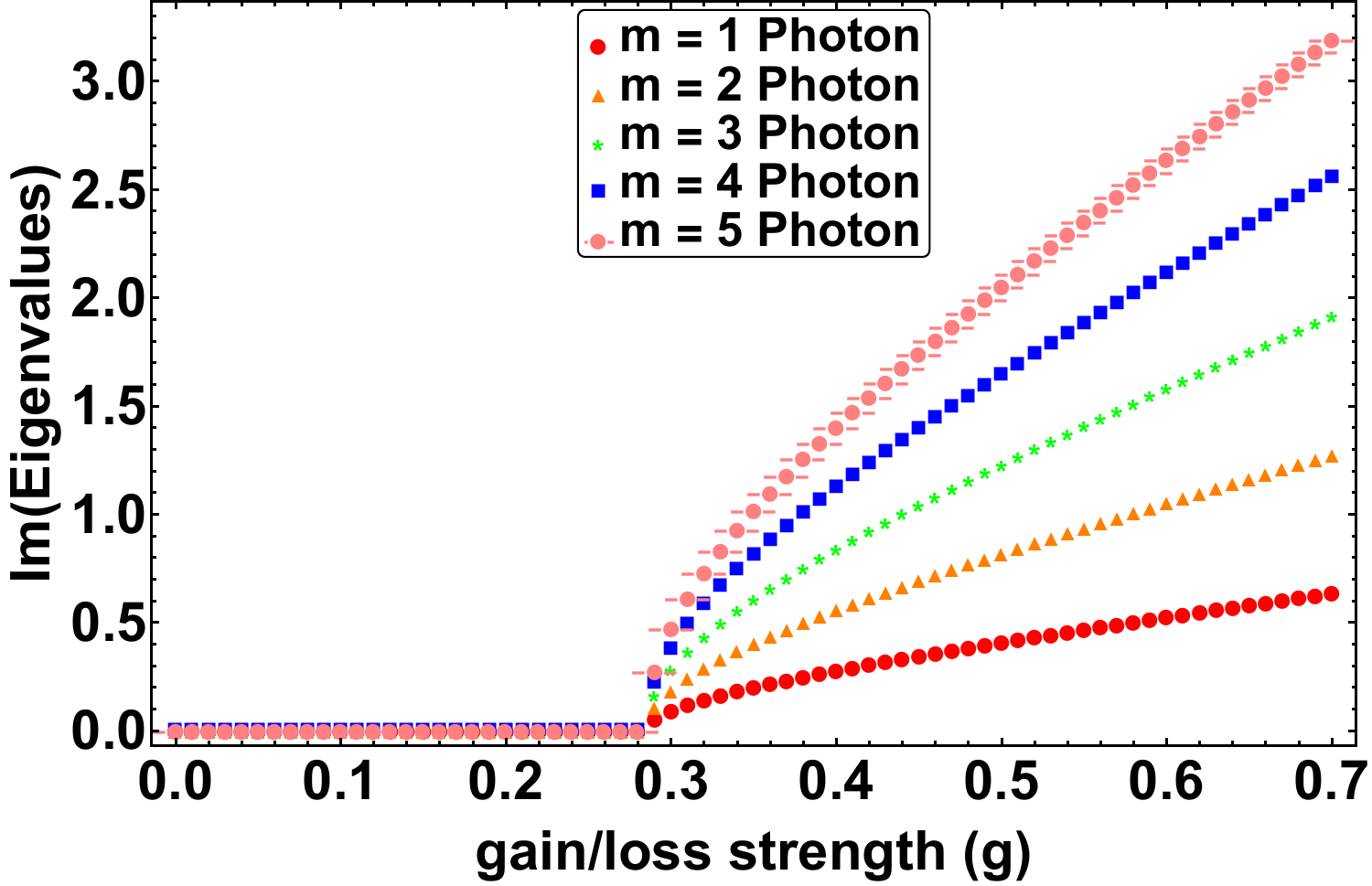}
		\caption{
			Imaginary parts of the eigenvalues as a function of the gain/loss parameter $g$ for a waveguide array with $N = 10$, shown for different photon-number subspaces.
		}
		\label{fig:fig10}
	\end{figure}
	Thus, based on the calculated propagation patterns for different quantum input states, the EP has been determined using the previously introduced criterion $I_r$. Figure~(\ref{fig:fig8}) illustrates the variation of $I_r$ with respect to the gain/loss coefficient $g$, for an entangled input state containing one photon distributed across the 24th and 25th waveguides in an array of $N = 10$ waveguides. \\
	The EP in this case is the transition point from one to zero in $I_r$ which is $g_c \simeq 0.289$ for this waveguide array. Through numerical analysis, it has been confirmed that the EP remains invariant with respect to the choice of input state; That is, the position of the EP remains unchanged for different quantum input states.
	This is essential because $g_c$ is a characteristic of the system itself and its corresponding Hamiltonian, not its input or initial state.
	\\
	The EP is typically identified as the critical value $g_c$ at which the system's eigenvalues become complex. As described in the theoretical section, this is done by constructing the reduced Hamiltonian in the $m$-photon subspace and computing its eigenvalues. The EP is then located by plotting the maximum imaginary part of the eigenvalues versus the gain/loss parameter $g$. Figure~(\ref{fig:fig9}) shows the maximum imaginary part of all eigenvalues as a function of $g$ for a waveguide array with $N = 10$ in the single-photon subspace ($m = 1$).
	
	Based on the plot, the EP for this case is determined to be $g_c = 0.284$, with a difference of only 0.005 compared to the EP obtained using $I_r$ (Figure~(\ref{fig:fig8})). This result shows the close values obtained by both methods.\\
	The maximum imaginary part of the eigenvalues, as observed for a waveguide array with $N = 10$ waveguides, in subspaces including one to five photons, is presented in figure~(\ref{fig:fig10}). As shown in figure~(\ref{fig:fig10}), although the maximum value of the imaginary parts of the eigenvalues differs across various photon subspaces, the onset of complex eigenvalues does not depend on the photon number $m$ in the subspace. In other words, increasing the number of photons does not affect the value of the EP. Therefore, it can be concluded that the method of calculating the EP using the reduced Hamiltonian in the $m$-photon subspace provides a general approach, and the EP of the full Hamiltonian is well-obtained.
	\\
	Afterwards, in addition to comparing the exceptional point $g_c$ determined by the two methods, the influence of the number of waveguides on the EP is investigated. Since the PT-symmetry pattern differs between arrays with even and odd numbers of waveguides—where even-numbered arrays exhibit alternating gain and loss, while in odd-numbered arrays the central waveguide is neutral and the rest alternate in gain and loss, the dependence of the EP on the number of waveguides is investigated separately for the even and odd cases.
	
	\begin{figure}[!t]
		\centering
		\begin{subfigure}{\linewidth}
			\centering
			\includegraphics[width=1\linewidth]{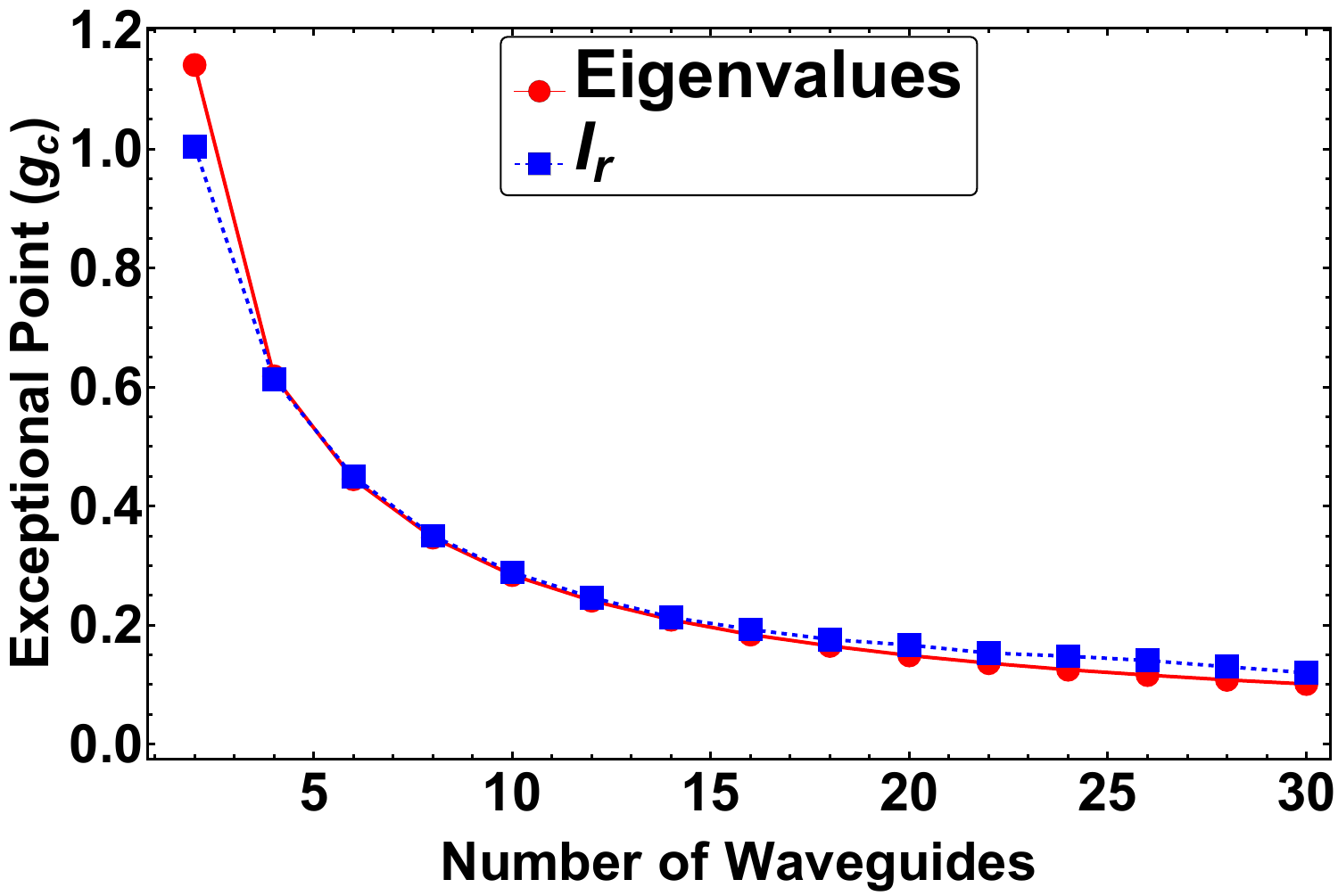}  
			\caption{}
			\label{fig:f11a}
		\end{subfigure}
		
		\vspace{0.5cm} 
		
		\begin{subfigure}{\linewidth}
			\centering
			\includegraphics[width=\linewidth]{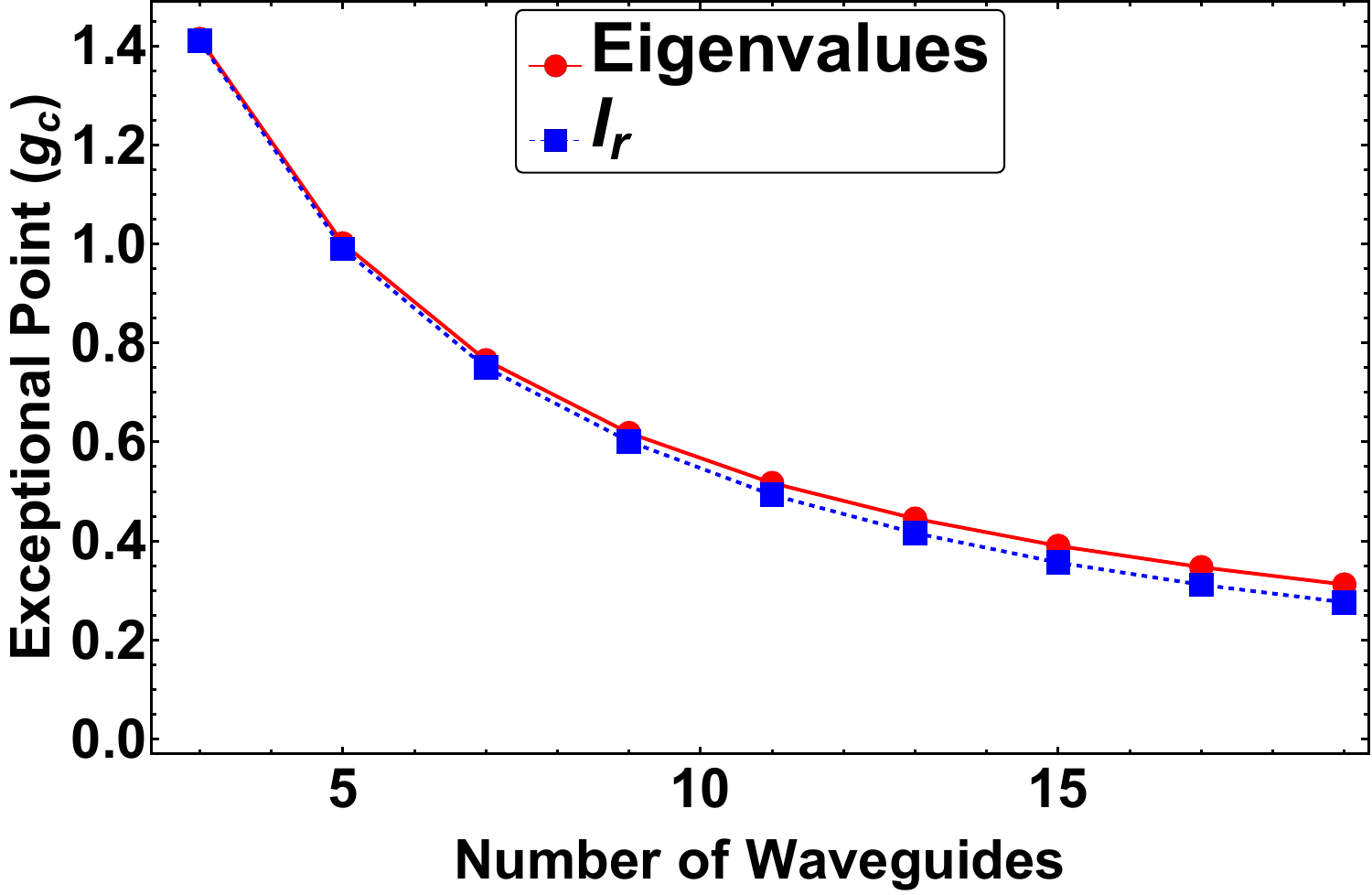}  
			\caption{}
			\label{fig:f11b}
		\end{subfigure}
		
		\caption{Comparison of the EP values obtained using both methods for (a) even and (b) odd numbers of waveguides.
		}
		\label{fig:fig11}
	\end{figure}
	As shown in Figure~(\ref{fig:fig11}), an increase in the number of waveguides leads to a decrease in $g_c$, indicating that the system enters the broken $\mathcal{PT}$-symmetric regime at lower gain/loss values. Furthermore, Figure~(\ref{fig:fig11}) confirms the consistency of the two approaches for identifying the $\mathcal{PT}$-symmetry breaking threshold. Another notable observation is that the value of the EP in arrays with an odd number of waveguides is higher than in those with an even number. This may be attributed to the presence of a central waveguide without gain or loss in arrays with an odd number of waveguides. The consistency between the EPs obtained via the two approaches is considered highly important. Given that the diagonalization of the Hamiltonian in PT-symmetric systems—especially in high-dimensional Hilbert spaces—is typically time-consuming, the $I_r$-based approach, which utilizes the propagation pattern, can serve as an effective alternative to the direct computation of eigenvalues.
	\label{sec3}
	
	\section{Summary and Conclusion}
	The propagation patterns in Hermitian and non-Hermitian (PT-symmetric) systems have been analyzed for number states, coherent states, squeezed states, and entangled states, and it was found that entangled states exhibit a propagation behavior distinct from the other types of quantum states. Furthermore, the site-to-site correlation function was investigated for different quantum input states. The second-order correlation function indicates the probability of simultaneously detecting two photons in distinct waveguides. The results indicate that, in the unbroken PT-symmetric regime, the statistical properties associated with photon intensity and site-to-site correlations remain stationary due to the balanced gain and loss, resembling those of a Hermitian system without gain or loss. One of the most notable differences in the site-to-site correlation pattern in the broken PT-symmetric phase is the concentration of the correlation function around the input waveguide. In other words, in this regime, photons are most likely to be detected in the gain-enhanced waveguides near the injection site.\\
	In addition, the EP was determined using two distinct methods. The results demonstrated consistency between the two approaches. Although computing the eigenvalues of the system is the direct approach for determining the EP, the strong agreement between both methods suggests that the $I_r$ criterion, which is based on the propagation pattern, can serve as an effective alternative for identifying the PT-symmetry breaking point. 
	For systems with exceptionally large Hilbert space dimensions, Hamiltonian diagonalization (eigenvalue-based methods~\cite{mailybaev2001evaluation}) becomes computationally infeasible. This is where the significance of the criterion introduced in this paper becomes apparent, as it can be calculated directly by solving differential equations related to intensity dynamics, without the need for diagonalization. This is particularly critical for complex quantum inputs such as coherent or squeezed states, where the total number of photons is not precisely defined, leading to unmanageable Hilbert space dimensions.
	Finally, the variation of the EP with respect to the number of waveguides was investigated, revealing that the EP value decreases as the number of waveguides in the array increases. In other words, the system enters the broken $\mathcal{PT}$-symmetric phase at a lower gain/loss threshold.
	\label{sec4}
	
    \nocite{*}  
    
     \bibliographystyle{ieeetr}  %
    \bibliography{ref}
    
    \section*{Author contributions statement}
    \textbf{Mahla Bahar:} Original draft writing, Data curation, Methodology, Investigation, Software.
    \textbf{Mojtaba Golshani:} Conceptualization, Methodology, Investigation, Software, Project administration, Writing review and editing.
    \textbf{Mostafa Motamedifar:} Formal analysis, Investigation, Writing review and editing.
    \textbf{Khatereh Jafari:} Investigation, Writing review and editing.

    \section*{Additional information}
    \textbf{Competing interests} The authors declare that they have no known competing financial interests or personal relationships that could have appeared to influence the work reported in this paper.
    
    \section*{Funding Statement}
    This research received no specific grant from any funding agency in the public, commercial, or not-for-profit sectors.
    
    \section*{Data Availability}
    The datasets used and/or analysed during the current study are available from the corresponding author on reasonable request.

	
\end{document}